\documentclass[useAMS,usenatbib]{mn2e}
\usepackage{amssymb}
\usepackage{amsmath}
\usepackage{graphicx}

\title[]{Application of screened Coulomb potential in fitting DBV star PG 0112+104}
\author[Y. H. Chen]{Y. H. Chen$^{1,2,3}$\thanks{E-mail: yhc1987@cxtc.edu.cn}\\
$^{1}$Institute of Astrophysics, Chuxiong Normal University, Chuxiong 675000, China\\
$^{2}$School of Physics and Electronical Science, Chuxiong Normal University, Chuxiong 675000,China\\
$^{3}$Key Laboratory for the Structure and Evolution of Celestial Objects, Chinese Academy of Sciences, P.O. Box 110, Kunming 650011, China\\}

\begin{document}

\date{Accepted: }

\pagerange{\pageref{firstpage}--\pageref{lastpage}} \pubyear{????}

\maketitle

\label{firstpage}

\begin{abstract}

With 78.7 days of observations for PG 0112+104, a pulsating DB star, from Campaign 8 of $Kepler$ 2 mission, Hermes et al. made a detailed mode identification. A reliable mode identification, with 5 $l$ = 1 modes, 3 $l$ = 2 modes, and 3 $l$ = 1 or 2 modes, was identified. Grids of DBV star models are evolved by \texttt{WDEC} with element diffusion effect of pure Coulomb potential and screened Coulomb potential. Fitting the identified modes of PG 0112+104 by the calculated ones, we studied the difference of element diffusion effect between adopting pure Coulomb potential and screened Coulomb potential. Our aim is to
reduce the fitting error by studying new input physics. The starting models including their chemical composition profile are from white dwarf models evolved by \texttt{MESA}. They were calculated following the stellar evolution from the main sequence to the start of the white dwarf cooling sequences. The optimal parameters are basically consistent with that of previous spectroscopic and asteroseismological studies. The pure and screened Coulomb potential lead to different composition profiles of the C/O-He interface area. High $k$ modes are very sensitive to the area. However, most of the observed modes for PG 0112+104 are low $k$ modes. The $\sigma_{RMS}$ taking the screened Coulomb potential is reduced by 4\% compared with taking the pure Coulomb potential when fitting the identified low $k$ modes of PG 0112+104. Fitting the $Kepler$ 2 data with our models improved the $\sigma_{RMS}$ of the fit by 27\%.

\end{abstract}


\section{Introduction}

White dwarfs are the final evolutionary state for most of low and medium mass stars, which corresponds to around 98\% of the end state of all stars (Winget \& Kepler 2008). For most white dwarfs, the nuclear fusion has basically stopped. Therefore, the white dwarfs gradually cool down through radiating. Along the evolution track of white dwarfs in the Hertzsprung-Russel diagram, there are DOV, DBV, and DAV instability strips. The DO type white dwarfs have rich helium (He) atmosphere with strong He II lines and the DOV instability strip is basically from 170000\,K to 75000\,K (Corsico et al. 2006). The DB type white dwarfs have rich He atmosphere with strong He I lines and the DBV instability strip is basically from 29000\,K to 22000\,K (Beauchamp et al. 1999, Bischoff-Kim \& {\O}stensen 2011). The DA type white dwarfs have rich hydrogen (H) atmosphere with only Balmer lines and the DAV instability strip is basically from 12270\,K to 10850\,K (Gianninas et al. 2005, Gianninas et al. 2011).

There are now at least 20 DBV stars observed (Corsico 2009). The DBV stars can be used as an excellent probe to study the energy loss rate for plasma neutrino. The neutrino luminosity and photon luminosity have different characteristics at the blue and red ends of the DBV instability strip (Winget et al. 2004). The rate of period change for hot DBV stars, such as the DBV star EC 20058-5234 and PG 0112+104, would be closely related to the plasma reaction. With pure-He DB models, Beauchamp et al. (1999) fitted EC 20058-5234 with effective temperature of $T_{eff}$ = 28400\,K and fitted PG 0112+104 with $T_{eff}$ = 31500\,K.

EC 20058-5234 is the eighth DBV star found with a magnitude of V=15.6 (Koen et al. 1995). Sullivan et al. (2008) presented an analysis of a total of 177\,h of high-quality optical time-series photometry on EC20058-5234 and identified 8 independent $l$ = 1 or 2, and $m$ = 0 modes. The index $l$ is the spherical harmonic degree and $m$ is the azimuthal number. Based on those 8 identified modes, they did asteroseismological study on EC 20058-5234 and then obtained optimal models with $T_{eff}$ $\sim$ 28200\,K. Their asteroseismological results are consistent with the pure He spectral fitting of $T_{eff}$ = 28400\,K (Beauchamp et al. 1999). The asteroseismology method is feasible and effective for the hot DBV star.

PG 0112+104 was first identified as a DB type white dwarf with strong He I lines by Greenstein et al. (1977). PG 0112+104 locates around the blue edge of the DBV instability strip. The photometric variability of PG 0112+104 was studied by Robinson \& Winget (1983), Kawaler et al. (1994), Shipman et al. (2002), and Dufour et al. (2010) subsequently. The amplitude of pulsations for PG 0221+104 are too low to be detected. With 78.7 d (duty cycle of 96.0\%) photometry from the $Kepler$ space telescope, Hermes et al. (2017) detected 11 independent pulsation modes with 5 low-order dipole modes and 3 quadrupole modes for PG 0112+104. PG 0112+104 indeed pulsates and becomes the hottest known DBV star. The mode identifications for PG 0112+104, based on rotational frequency splitting, are quite reliable. The identified $m$ = 0 modes can be used to constrain fitting models and the frequency splitting values can be used to probe the internal rotation at different depths.

DB type stars have ideal gas of He atmosphere on the degenerate carbon/oxygen (C/O) core. They have stellar masses around 0.60\,$M_{\odot}$ (Voss et al. 2007, Bergeron et al. 2011). The stellar evolution theory constrains log($M_{\rm He}/M_{\rm *}$) to the range -2.0 to -6.0, where $M_{\rm He}$ is the helium layer mass and $M_{\rm *}$ is the stellar mass. If log($M_{\rm He}/M_{\rm *}$) is larger than -2.0, the helium will be ignited (Iben 1991) and if log($M_{\rm He}/M_{\rm *}$) is smaller than -6.0, the helium atmosphere will be polluted by carbon (Pelletier et al. 1986). The effects of element diffusion and gravitational sedimentation make the white dwarfs stratification structure. The element diffusion scheme is important for the composition profiles and the detailed pulsation periods. Thoul et al. (1994) reported an element diffusion scheme in the solar interior. Adding the element diffusion scheme into the White Dwarf Evolution Code (\texttt{WDEC}), the element diffusion effect for white dwarfs was studied by Su et al. (2014). The expression for the Coulomb logarithm (Eq.\,(9) in their paper) is for a pure Coulomb potential with a cutoff at the Debye radius. They use an analytical fit of Iben \& MacDonald (1985) obtained from the results of Fontaine \& Michaud (1979). However, Paquette et al. (1986) reported that a screened Coulomb potential would provide a better description of the plasma for white dwarfs.

The mode identifications for PG 0112+104 based on rotational frequency splitting are quite reliable for DBV stars (Hermes et al. 2017). The observed data, especially for the data set from the $Kepler$ space telescope, usually has very high precision. The model fitting quality, in this field, is far worse than the quality of the observed data. As an additional motivation, studying the element diffusion effect of pure Coulomb potential and screened Coulomb potential is a good attempt of improving the model input physics. We try to do an asteroseismological study on PG 0122+104 taking the element diffusion effect of pure Coulomb potential and screened Coulomb potential into account. We can compare the fitting results with element diffusion effect of pure Coulomb potential and screened Coulomb potential. In Sect. 2, the photometric observations and mode identifications for PG 0112+104 are briefly reviewed. In Sect. 3, we introduce the input physics and model calculations. The asteroseismological study on PG 0112+104 are showed in Sect. 4. In Sect. 5, we give a discussion and conclusions.

\section{Photometric observations and mode identifications for PG 0112+104}

Provencal et al. (2003) reported two low amplitude pulsations at 197.76$\pm$0.01 (0.874$\pm$0.14 mma) and 168.97$\pm$0.01 (0.833$\pm$0.14 mma) seconds for PG 0112+104, based on 30 hours of white-light photometry. The two periods are consistent with low-nodes $g$-mode pulsations for DBV stars. During Campaign 8 of the $Kepler$ 2 mission, PG 0112+104 was observed for 78.7 days from 2016 January 04. Hermes et al. (2017) made a discrete Fourier transform on the released data and then they identified 11 independent modes, as shown in Table 1 (from Table 1 of Hermes et al. (2017)). The two low amplitude modes identified by Provencal et al. (2003) were reproduced as $f_{1a}$ and $f_{2b}$ in Table 1.

\begin{table}
\begin{center}
\begin{tabular}{lcccccccccccccc}
\hline
ID          &$Fre.$         &$Per.$     &$Amp.$       &$\delta$$Fre.$     \\
            &($\mu$Hz)      &(s)        &(ppt)        &($\mu$Hz)          \\
\hline
$f_{1a}$    & 5055.9298(52) & 197.78756 & 0.298 &                         \\
            &               &           &       &14.8659                  \\
$f_{1b}$    & 5070.7957(56) & 197.20771 & 0.281 &                         \\
            &               &           &       &14.8653                  \\
$f_{1c}$    & 5085.6610(59) & 196.63127 & 0.267 &                         \\
\hline
$f_{2a}$    & 5893.4822(99) & 169.67897 & 0.157 &                         \\
            &               &           &       &23.8373                  \\
$f_{2b}$    & 5917.3195(63) & 168.99544 & 0.249 &                         \\
            &               &           &       &47.5745                  \\
$f_{2d}$    & 5964.8940(56) & 167.64757 & 0.281 &                         \\
            &               &           &       &23.7166                  \\
$f_{2e}$    & 5988.6106(87) & 166.98364 & 0.179 &                         \\
\hline
$f_{3a}$    & 3614.4976(56) & 276.66362 & 0.277 &                         \\
            &               &           &       &14.4914                  \\
$f_{3b}$    & 3628.9890(88) & 275.55884 & 0.177 &                         \\
            &               &           &       &14.5092                  \\
$f_{3c}$    & 3643.4982(56) & 274.46150 & 0.278 &                         \\
\hline
$f_{4a?}$   & 4054.2233(87) & 246.65637 & 0.179 &                         \\
            &               &           &       &15.3996                  \\
$f_{4b?}$   & 4069.6229(57) & 245.72301 & 0.275 &                         \\
\hline
$f_{5a}$    & 6586.1032(95) & 151.83485 & 0.165 &                         \\
            &               &           &       &23.7225                  \\
$f_{5b}$    & 6609.8257(83) & 151.28992 & 0.188 &                         \\
            &               &           &       &47.3101                  \\
$f_{5d}$    & 6657.1358(70) & 150.21475 & 0.223 &                         \\
            &               &           &       &23.6132                  \\
$f_{5e}$    & 6680.749(12)  & 149.68382 & 0.127 &                         \\
\hline
$f_{6d?}$   & 5155.0647(71) & 193.98399 & 0.221 &                         \\
            &               &           &       &23.1521                  \\
$f_{6e?}$   & 5178.2168(88) & 193.11667 & 0.178 &                         \\
\hline
$f_{7b?}$   & 3129.7151(71) & 319.51790 & 0.221 &                         \\
            &               &           &       &16.1769                  \\
$f_{7c?}$   & 3145.892(14)  & 317.8748  & 0.111 &                         \\
\hline
$f_{8b?}$   & 2801.2608(74) & 356.98212 & 0.212 &                         \\
\hline
$f_{9b/d?}$ & 2011.357(11)  & 497.1768  & 0.140 &                         \\
\hline
$f_{10d?}$  & 4643.606(12)  & 215.34986 & 0.131 &                         \\
\hline
$f_{11b?}$  & 6277.488(15)  & 159.29938 & 0.105 &                         \\
            &               &           &       &17.726                   \\
$f_{11c?}$  & 6295.214(18)  & 158.85084 & 0.086 &                         \\
\hline
\end{tabular}
\caption{Mode identifications for PG 0122+104 by Hermes et al. (2017). In the header, $Fre.$ is pulsation frequency, $Per.$ is corresponding period, $Amp.$ is the amplitude, and $\delta$$Fre.$ is a value of frequency difference.}
\end{center}
\end{table}

The modes with frequency splitting values basically from 14.5\,$\mu$Hz to 17.7\,$\mu$Hz were identified as $l$ = 1 modes. Modes of $f_{1}$ and $f_{3}$ show complete triplets, while modes of $f_{4}$, $f_{7}$, and $f_{11}$ show doublets. The modes of $f_{2}$, $f_{5}$, and $f_{6}$ have frequency splitting around 23.5\,$\mu$Hz. They were identified as $l$ = 2 modes, with incomplete components. Brickhill (1975) derived an approximate formula between frequency splitting ($\delta\nu_{n,l}$) and rotational period ($P_{\rm rot}$) as
\begin{equation}
m\delta\nu_{n,l}=\nu_{n,l,m}-\nu_{n,l,0}=\frac{m}{P_{\rm rot}}(1-\frac{1}{l(l+1)}),
\end{equation}
\noindent where $n$ is the radial order. According to Eq.\,(1), $\delta\nu_{n,1}$/$\delta\nu_{n,2}$ is 0.6. The value of 14.5\,$\mu$Hz/23.5\,$\mu$Hz is basically 0.6. In Table 1, $f_{1b}$ and $f_{3b}$ is the $m$ = 0 components in their triplets. The modes of $f_{2c}$ and $f_{5c}$ are not observed and they are the $m$ = 0 components in their quintuplets. For $l$ = 1 modes, we mark a subscript $b$ as the $m$ = 0 component. For $l$ = 2 modes, we mark a subscript $c$ as the $m$ = 0 component. The amplitude relations for modes of $f_{2}$, $f_{3}$, and $f_{5}$ show that the angle of inclination of rotation axis ($i$) seems around $60^{\circ}$ (Pesnell 1985). Therefore, Hermes et al. (2017) identified the observed $f_{6}$ modes as $f_{6d}$ and $f_{6e}$ (the amplitude of $f_{6d}$ is larger than that of $f_{6e}$). They calculated $f_{6c}$ ($m$ = 0) as 194.8591\,s. The amplitudes of $f_{2d}$ and $f_{5d}$ are larger than other components. The mode of $f_{10}$ is close to the mode of $f_{1}$. Therefore, it was identified as $f_{10d}$ corresponding to $f_{10c}$ of 216.447\,s. However, for $l$ = 1 modes, the amplitudes of $f_{1}$ modes are not consistent with the geometric effect of $i$ $\sim$ $60^{\circ}$. In fact, the amplitude of $f_{1a}$ ($f_{2b}$) is various from 0.874 mma (0.833 mma) (Provencal et al. 2003) to 0.298 mma (0.249 mma) (Hermes et al. 2017). For simplicity, the mode with large amplitude was identified as the $m$ = 0 component for $l$ = 1 doublets. Therefore, $f_{4a}$, $f_{4b}$, $f_{7b}$, $f_{7c}$, $f_{11b}$, $f_{11c}$, and $f_{8b}$ were identified. The mode of $f_{9}$ was identified as $f_{9b}$ ($l$ = 1) or $f_{9d}$ ($l$ = 2 and $f_{9c}$ is 503.062\,s). The detailed $m$ = 0 modes are showed in Table 2.

\begin{table}
\begin{center}
\begin{tabular}{lcccccccccccccc}
\hline
ID          &$Per.$               &$l$           \\
            &(s)                  &              \\
\hline
$f_{11b?}$  & 159.29938           &1             \\
$f_{1b}$    & 197.20771           &1             \\
$f_{4b?}$   & 245.72301           &1             \\
$f_{3b}$    & 275.55884           &1             \\
$f_{7b?}$   & 319.51790           &1             \\
$f_{8b?}$   & 356.98212           &1or2          \\
$f_{5c}$    & 150.7506            &2             \\
$f_{2c}$    & 168.3185            &2             \\
$f_{6c?}$   & 194.8591            &2             \\
$f_{10c?}$  & 216.447             &2or1          \\
$f_{9b/d?}$ & 497.1768/503.062    &1/2          \\
\hline
\end{tabular}
\caption{Mode identifications for PG 0122+104 by Hermes et al. (2017).}
\end{center}
\end{table}

With the identified modes in Table 2, Hermes et al. (2017) tried to study the differential rotational effect. They evolved grids of DBV star models by an updated version of \texttt{WDEC}. They adopted the parameterized structure without element diffusion (Hermes et al. 2017, Bischoff-Kim et al. 2014, Metcalfe 2015). The evolved DBV stars were used to fit the 11 modes in Table 2. However, the root-mean-square residual $\sigma_{RMS}$ is 2.471\,s for their best-fitting model. They also performed a surface rotation rate of 10.17404 hr from an apparent spot modulation in the $Kepler$ 2 data. For the observed data, according to Eq.\,(1), 14.5\,$\mu$Hz (17.7\,$\mu$Hz) for $l$ = 1 modes corresponds to a rotation rate of 9.58 hr (7.85 hr) and 23.2\,$\mu$Hz (23.8\,$\mu$Hz) for $l$ = 2 modes corresponds to a rotation rate of 9.98 hr (9.73 hr). However, the present asteroseismology model with $\sigma_{RMS}$ = 2.471\,s is not good enough to probe the differential rotation with depths. In our work, the modes identified in Table 2 are used to check the DBV star models with the application of screened Coulomb potential. The input physics and model calculations are displayed in the next section.

\section{Input physics and model calculations}

In this section, we introduce the input physics and model calculations. Paxton et al. (2011, 2013) reported a stellar evolution code of Modules for Experiments in Stellar Astrophysics (\texttt{MESA}). Based on thermal nuclear reactions (Caughlan \& Fowler 1988, Angulo et al. 1999), \texttt{MESA} can evolve a star from main-sequence ($MS$) stage to the white dwarf ($WD$) stage. The \texttt{MESA} version 6208 is downloaded and installed. With the module '$make\_co\_wd$', a group of $MS$ stars are evolved to the $WD$ stage, as shown in Table 3.

\begin{table}
\begin{center}
\begin{tabular}{lccccc}
\hline
$MS$         &$WD$ core(\texttt{MESA})   &$WD$ mass range(\texttt{WDEC})   \\
($M_{\odot}$)&($M_{\odot}$)              &($M_{\odot}$)                    \\
\hline
2.0          &0.475                      &0.460-0.485                      \\
2.5          &0.500                      &0.490-0.510                      \\
2.8          &0.525                      &0.515-0.535                      \\
3.0          &0.550                      &0.540-0.560                      \\
3.2          &0.575                      &0.565-0.585                      \\
3.4          &0.600                      &0.590-0.600                      \\
\hline
\end{tabular}
\caption{Masses of $MS$ progenitors with corresponding $WD$ core masses in \texttt{MESA} and approximate $WD$ mass range in \texttt{WDEC}.}
\end{center}
\end{table}

\texttt{WDEC} was first derived by Schwarzschild. It was modified and updated by Kutter \& Savedoff (1969), Lamb \& van Horn (1975), and Wood (1990) respectively. \texttt{WDEC} can evolve white dwarfs with artificially constructed C/O core compositions and C/O-He, He/H diffusive equilibrium profiles. Itoh et al. (1983, 1984) calculated the opacities. The equation of state is derived by Lamb (1974) and Saumon, Chabrier \& van Horn (1995). The standard mixing length theory is used (B\"{o}hm \& Cassinelli 1971, Tassoul, Fontaine \& Winget 1990). The mixing length parameter (mixing length over pressure scale height) $\alpha$ is adopted as 1.25 for calculations (Beauchamp et al. 1999, Bergeron et al. 2011). The $WD$ cores evolved by \texttt{MESA}, as shown in the second column in Table 3, are used to match the $WD$ stars evolved by \texttt{WDEC}. The $WD$ cores evolved by \texttt{MESA}, together with structure parameters of mass, radius, luminosity, pressure, temperature, entropy, and C abundance, are inserted into \texttt{WDEC} to evolve DBV stars. The corresponding $WD$ masses are shown in the third column in Table 3. Su et al. (2014) added the element diffusion scheme of Thoul et al. (1994) into \texttt{WDEC} and did an asteroseismological study on a DAV star KUV 11370+4222. The equation of Coulomb logarithm (Eq.\,(9) in Thoul et al. 1994) is treated as a pure Coulomb potential with a cutoff at the Debye radius. The DBV stars evolved in this method are treated as results of taking the element diffusion effect with pure Coulomb potential into account.

The scheme of Thoul et al. (1994) was used in the solar interior. For white dwarfs, a Debye-like potential of
\begin{equation}
V_{st}(x)=\frac{Z_{s}Z_{t}e^{2}}{r}e^{-(r/\lambda)}
\end{equation}
\noindent is used. In Eq.\,(2), $Z_{s}$, $Z_{t}$ are particle charges and $r$ is the distance. The parameter $\lambda$ is the greater one between the Debye length $\lambda_{D}$ and the mean ionic distance, $a_{0}$ = $[3/4\pi n_{i}]^{1/3}$, where $n_{i}$ is the ionic number density.

The Burgers equations for momentum and energy conservation (Eq.\,(12) and (13) in Thoul et al. 1994) are replaced by (Muchmore 1984, Cox, Guzik \& Kidman 1989)
\begin{equation}
\begin{split}
  &\nabla p_{s}-\frac{\rho_{s}}{\rho}\nabla p-n_{s}q_{s}E=\\
  &\sum_{t}K_{st}(\omega_{t}-\omega_{s})+\sum_{t}K_{st}z_{st}\frac{m_{t}r_{s}-m_{s}r_{t}}{m_{s}+m_{t}},
\end{split}
\end{equation}
\noindent and
\begin{equation}
\begin{split}
   &\frac{5}{2}n_{s}k\nabla T=\\
  -&\frac{5}{2}\sum_{t\neq s}K_{st}z_{st} \frac{m_{t}}{m_{s}+m_{t}} (\omega_{t}-\omega_{s})-\frac{2}{5}K_{ss}z''_{st}r_{s}\\
  -&\sum_{t\neq s}\frac{K_{st}}{(m_{s}+m_{t})^{2}}(3m_{s}^{2}+m_{t}^{2}z'_{st}+0.8m_{s}m_{t}z''_{st})r_{s}\\
  +&\sum_{t\neq s}\frac{K_{st}m_{s}m_{t}}{(m_{s}+m_{t})^{2}}(3+z'_{st}-0.8z''_{st})r_{t}.
\end{split}
\end{equation}
\noindent In Eq.\,(3) and (4), $p_{s}$, $\rho_{s}$, $n_{s}$, $q_{s}$, and $m_{s}$ are respectively the partial pressure, mass density, number density, charge, and mass for species $s$. The parameter $\omega$ is the diffusion velocity, $r$ is the residual heat flow vector, $T$ is the temperature, and $E$ is the electric field. An independent variable $\phi$ (Eq.\,(21) in Muchmore 1984) is defined as
\begin{equation}
\phi=ln[ln(1+\gamma_{st}^{2})],
\end{equation}
\noindent where,
\begin{equation}
\gamma_{st}=\frac{4kT \lambda}{Z_{s} Z_{t} e^{2}}.
\end{equation}
\noindent
Muchmore (1984) calculated the resistance coefficients $K_{st}$, $z_{st}$, and $z'_{st}$ by polynomial fittings and made $z''_{st}$ approximately equal to 2.62. According to Eq.\,(3) in Iben \& Macdonald (1985), the Eq.\,(9) in Thoul et al. (1994) is expressed as a function of Eq.\,(22) in Muchmore (1984),
\begin{equation}
\begin{split}
ln \Lambda_{st}=&0.5ln(1+\gamma_{st}^{2})(0.36718-1.6996\times10^{-2}\phi\\
                &+5.7908\times10^{-2}\phi^{2}+2.4384\times10^{-3}\phi^{3}).
\end{split}
\end{equation}
\noindent By replacing Eq.\, (9, 12, 13) in Thoul et al. (1994) with Eq.\,(1, 2, 22-25) in Muchmore (1984), the evolved DBV stars are treated as results of taking the element diffusion effect with screened Coulomb potential into account. According to Fig. 3 of Zhang (2017), the calculated potential in this method is the repulsive potential. We will test the effect of screened Coulomb potential by an analysis of the pulsation spectrum of PG 0112+104.

Calculating the diffusion effect with pure Coulomb potential and screened Coulomb potential, two grids of DBV star models are evolved. Dufour et al. (2010) derived a self-consistent $T_{eff}$ of 31,300$\pm$500 K, a surface gravity of log$g$ = 7.8$\pm$0.1 ($M_{\rm *}$ = 0.52$\pm$0.05\,$M_{\odot}$) for PG 0112+104, based on optical data. Therefore, we just evolve $MS$ stars of 2.0\,$M_{\odot}$-3.4\,$M_{\odot}$, which corresponds to $WD$ core masses of 0.475\,$M_{\odot}$-0.600\,$M_{\odot}$. The effective temperature $T_{\rm eff}$ is from 34000\,K to 28000\,K with a step of 200\,K. The total stellar mass $M_{\rm *}$ is from 0.460\,$M_{\odot}$ to 0.600\,$M_{\odot}$ with a step of 0.005\,$M_{\odot}$. The helium mass fraction log($M_{\rm He}/M_{\rm *}$) is from -4.0 to -7.0 with a step of 0.5. To evolve DBV star models, the hydrogen mass fraction is set as infinitely small (log($M_{\rm H}/M_{\rm *}$) = -200) and the hydrogen abundance is set as 0.0. For each grid of DBV star models, more than 6000 models are evolved. Then, based on the pulsation code of Li (1992a,b), we numerically solve the full equations of linear and adiabatic oscillation on those DBV star models. The eigen-frequencies can be found one by one and then be used to fit the observed modes in Table 2.

\section{The asteroseismological study on PG 0112+104}

In Sect. 2, we review the mode identifications for PG 0112+104. There are 5 $l$ = 1 modes, 3 $l$ = 2 modes, 3 $l$ = 1 or 2 modes identified, as shown in Table 2. The 11 modes are used to constrain the fitting models. In Sect. 3, we introduce the method of evolving DBV star models taking the element diffusion effect with pure and screened Coulomb potential into account. In this section, we discuss the asteroseismological study on PG 0112+104. The grids of DBV star models are used to fit the observed modes in Table 2. The root-mean-square residual ($\sigma_{RMS}$) is used to evaluate the quality of the fitting results, which is expressed by
\begin{equation}
\sigma_{RMS}=\sqrt{\frac{1}{n} \sum_{n}(Per.-P_{\rm mod})^2}.
\end{equation}
\noindent In Eq.\,(8), $n$ is the number of observed modes. It is 11 for PG 0112+104 in this paper. The smaller the value of $\sigma_{RMS}$, basically for a fitting model, the better the fitting results.

\begin{figure}
\begin{center}
\includegraphics[width=9.0cm,angle=0]{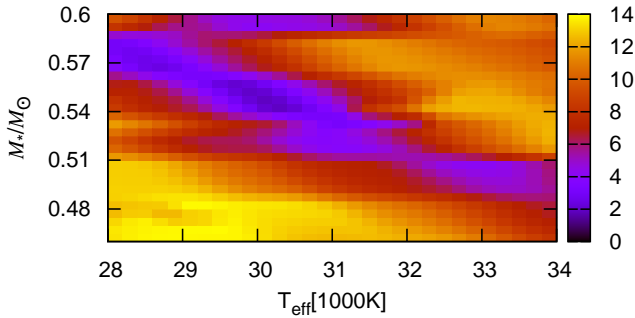}
\end{center}
\caption{The colour residual diagram for fitting PG 0112+104 with DBV star models adopting the screened Coulomb potential. For the models, log($M_{\rm He}/M_{\rm *}$) is -6.0. The colours indicate the values of $\sigma_{RMS}$. For the optimal model, $M_{\rm *}$ = 0.545\,$M_{\odot}$, $T_{\rm eff}$ = 30200\,K, and $\sigma_{RMS}$ = 1.806\,s.}
\end{figure}

\begin{figure}
\begin{center}
\includegraphics[width=9.0cm,angle=0]{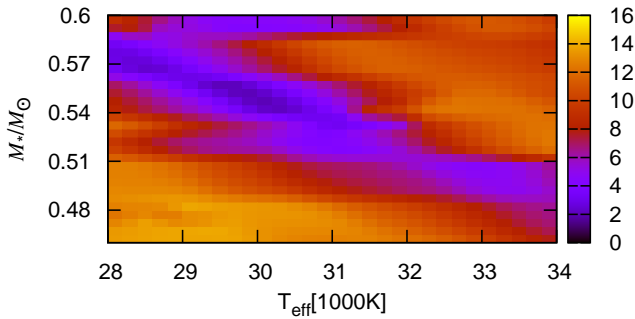}
\end{center}
\caption{The colour residual diagram for fitting PG 0112+104 with DBV star models adopting the pure Coulomb potential. For the models, log($M_{\rm He}/M_{\rm *}$) is -6.0. The colours indicate the values of $\sigma_{RMS}$. For the optimal model, $M_{\rm *}$ = 0.550\,$M_{\odot}$, $T_{\rm eff}$ = 30000\,K, and $\sigma_{RMS}$ = 1.885\,s.}
\end{figure}

For the DBV star models with both screened Coulomb potential and pure Coulomb potential, the helium atmospheres for the optimal models are log($M_{\rm He}/M_{\rm *}$) = -6.0. The colour residual diagram is shown in Fig. 1 and Fig. 2 respectively. For the scenario of taking screened Coulomb potential, the optimal model is $M_{\rm *}$ = 0.545\,$M_{\odot}$, $T_{\rm eff}$ = 30200\,K, and $\sigma_{RMS}$ = 1.806\,s. For the scenario of taking pure Coulomb potential, the optimal model is $M_{\rm *}$ = 0.550\,$M_{\odot}$, $T_{\rm eff}$ = 30000\,K, and $\sigma_{RMS}$ = 1.885\,s. The two model parameters are very close, with only 0.005\,$M_{\odot}$ and 200\,K differences.

\begin{table*}
\begin{center}
\begin{tabular}{lllllllllllll}
\hline
$P_{\rm mod}(l,k)$&$Per.$&$\sigma_{p}$&$P_{\rm mod}(l,k)$&$Per.$&$\sigma_{p}$&$P_{\rm mod}(l,k)$&$Per.$&$\sigma_{p}$\\
(s)           &(s)      &(s)         &(s)           &(s)      &(s)         &(s)           &(s)      &(s)         \\
\hline
157.580(1,1)  &159.29938&$\,$ 1.719  &102.910(2,1)  &         &            &766.863(2,29) &         &            \\
193.845(1,2)  &197.20771&$\,$ 3.363  &118.596(2,2)  &         &            &793.337(2,30) &         &            \\
246.774(1,3)  &245.72301&$\,$-1.051  &149.410(2,3)  &150.7506 &$\,$ 1.341  &819.722(2,31) &         &            \\
275.600(1,4)  &275.55884&$\,$-0.041  &168.445(2,4)  &168.3185 &$\,$-0.126  &842.500(2,32) &         &            \\
319.153(1,5)  &319.51790&$\,$ 0.365  &191.621(2,5)  &194.8591 &$\,$ 3.238  &863.485(2,33) &         &            \\
358.804(1,6)  &356.98212&$\,$-1.822  &218.312(2,6)  &216.447  &$\,$-1.865  &892.804(2,34) &         &            \\
395.636(1,7)  &         &            &240.875(2,7)  &         &            &917.812(2,35) &         &            \\
442.684(1,8)  &         &            &264.848(2,8)  &         &            &941.374(2,36) &         &            \\
480.241(1,9)  &         &            &290.819(2,9)  &         &            &966.718(2,37) &         &            \\
526.279(1,10) &         &            &312.580(2,10) &         &            &990.077(2,38) &         &            \\
567.051(1,11) &         &            &338.000(2,11) &         &            &1019.332(2,39)&         &            \\
602.556(1,12) &         &            &360.839(2,12) &         &            &1042.709(2,40)&         &            \\
647.548(1,13) &         &            &385.047(2,13) &         &            &1067.045(2,41)&         &            \\
687.635(1,14) &         &            &411.078(2,14) &         &            &1095.744(2,42)&         &            \\
729.733(1,15) &         &            &430.078(2,15) &         &            &1120.061(2,43)&         &            \\
767.798(1,16) &         &            &453.498(2,16) &         &            &1146.801(2,44)&         &            \\
805.742(1,17) &         &            &478.996(2,17) &         &            &1170.119(2,45)&         &            \\
854.956(1,18) &         &            &504.184(2,18) &503.062  &$\,$-1.122  &1196.182(2,46)&         &            \\
892.402(1,19) &         &            &527.671(2,19) &         &            &1225.993(2,47)&         &            \\
928.160(1,20) &         &            &547.081(2,20) &         &            &1249.657(2,48)&         &            \\
973.210(1,21) &         &            &573.345(2,21) &         &            &1277.128(2,49)&         &            \\
1015.870(1,22)&         &            &599.976(2,22) &         &            &              &         &            \\
1062.469(1,23)&         &            &621.353(2,23) &         &            &              &         &            \\
1096.838(1,24)&         &            &643.932(2,24) &         &            &              &         &            \\
1133.149(1,25)&         &            &668.980(2,25) &         &            &              &         &            \\
1184.138(1,26)&         &            &695.796(2,26) &         &            &              &         &            \\
1226.828(1,27)&         &            &720.610(2,27) &         &            &              &         &            \\
1267.353(1,28)&         &            &739.835(2,28) &         &            &              &         &            \\
\hline
\end{tabular}
\caption{The fitting result of the optimal model adopting screened Coulomb potential. The value $\sigma_{p}$ is $Per.$ minus $P_{\rm mod}(l,k)$ in seconds.}
\end{center}
\end{table*}

\begin{table*}
\begin{center}
\begin{tabular}{lccccc}
\hline
ID   &$T_{\rm eff}$ &log\,$g$     &$M_{*}$      &log($M_{\rm He}/M_{*}$)&$\sigma_{RMS}$\\
     &(K)           &             &($M_{\odot}$)&                       &(s)           \\
\hline
1    &31300$\pm$500 &7.8$\pm$0.1  &0.52$\pm$0.05&                       &              \\
2    &31040$\pm$1060&7.83$\pm$0.06&             &                       &              \\
3    &30600         &             &0.54         &                       &2.471         \\
4(s) &30200         &             &0.545        &-6.0                   &1.806         \\
4(p) &30000         &             &0.550        &-6.0                   &1.885         \\
\hline
\end{tabular}
\caption{Table of optimal models. The ID number 1, 2, 3, and 4 means results of Dufour et al. (2010), Bergeron et al. (2011), Hermes et al. (2017), and the present paper, respectively.}
\end{center}
\end{table*}

In Table 4, we show the optimal model selected in Fig. 1 and the detailed fitting results. The calculated range of pulsation periods is from 100\,s to 1300\,s. There are 28 $l$ = 1 modes and 49 $l$ = 2 modes for the model of $M_{\rm *}$ = 0.545\,$M_{\odot}$, $T_{\rm eff}$ = 30200\,K, and log($M_{\rm He}/M_{\rm *}$) = -6.0. In table 2, totally 6 modes are fitted by $l$ = 1 modes and 5 modes are fitted by $l$ = 2 modes. For the model, the value of $\sigma_{RMS}$ is 1.806\,s. We also show a parameter $\sigma_{p}$, which is the observed period ($Per.$) minus the calculated period ($P_{\rm mod}(l,k)$). The maximum value of the absolute value of $\sigma_{p}$ is 3.363\,s, fitting the observed mode of $f_{1b}$. Except for fitting the modes of $f_{1b}$ and $f_{6c}$, the fitting errors of other modes are less than 2 seconds. The mode of $f_{9d}$ is identified as 503.062\,s, being fitted by 504.184\,s ($l$=2, $k$=18).

In Table 5, we show the optimal models of spectroscopic and asteroseismological results. The values of $T_{\rm eff}$ of two asteroseismological studies are slightly smaller than that of the spectroscopic study of Dufour et al. (2010) and consistent with that of the spectroscopic study of Bergeron et al. (2011). The values of $M_{*}$ of two asteroseismological studies are consistent with that of the spectroscopic study of Dufour et al. (2010).

Hermes et al. (2017) made grids of the central O abundance relative to C, the mass fraction of the homogeneous C/O core, the location of the mixed C/He layer, and the location of the He abundance rising to 1. They adopted the parameterized structure, without element diffusion (Hermes et al. 2017, Bischoff-Kim et al. 2014, Metcalfe 2015). Their optimal model has $\sigma_{RMS}$ = 2.471\,s. They identified the $l$ = 1 modes as $k$ = 2-7, the $l$ = 2 modes as $k$ = 4-7 except the relatively long-period mode $f_{9}$. We insert the core compositions of thermal nuclear burning results from white dwarf models evolved by \texttt{MESA} into \texttt{WDEC} to evolve DBV stars. The element diffusion scheme is added with adopting pure and screened Coulomb potential. For these methods, the optimal models are very close, as shown in Fig. 1 and Fig. 2. For the scenario of taking screened Coulomb potential, the optimal model has $\sigma_{RMS}$ = 1.806\,s, which is a substantial improvement on the previous analysis. In addition, we identify the $l$ = 1 modes as $k$ = 1-6, the $l$ = 2 modes as $k$ = 3-6, 18, as shown in Table 4. The major difference between the Hermes et al. (2017) and the present study is the use of detailed stellar models as input for the WDEC code.

\begin{figure}
\begin{center}
\includegraphics[width=9.0cm,angle=0]{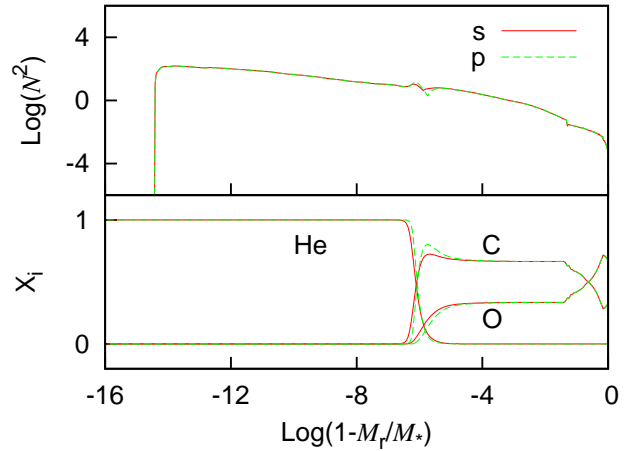}
\end{center}
\caption{The composition profiles and corresponding Brunt-V\"ais\"al\"a frequency for two models of $M_{*}$ = 0.545\,$M_{\odot}$, $T_{\rm eff}$ = 30200\,K, and log($M_{\rm He}/M_{*}$) = -6.0. The red solid line is the model evolved by adopting the screened Coulomb potential, while the green dashed one is the model evolved by adopting the pure Coulomb potential.}
\end{figure}

\begin{figure}
\begin{center}
\includegraphics[width=9.0cm,angle=0]{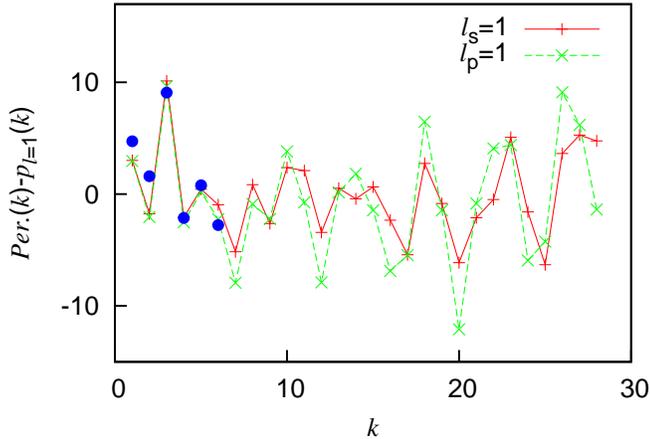}
\end{center}
\caption{The fitting diagram for $l$ = 1 modes. The two models are same with which in Fig. 3. The red one is the model evolved by adopting the screened Coulomb potential, while the green one is the model evolved by adopting the pure Coulomb potential. The blue filled dots represent the observed modes for $l$ = 1. The calculated $l$ = 1 modes in Table 4 are linear fitted. The figure is drawn by subtracting the fitting function from both the observed modes ($Per.$) and the calculated modes ($P_{\rm mod}(l,k)$).}
\end{figure}

\begin{figure}
\begin{center}
\includegraphics[width=9.0cm,angle=0]{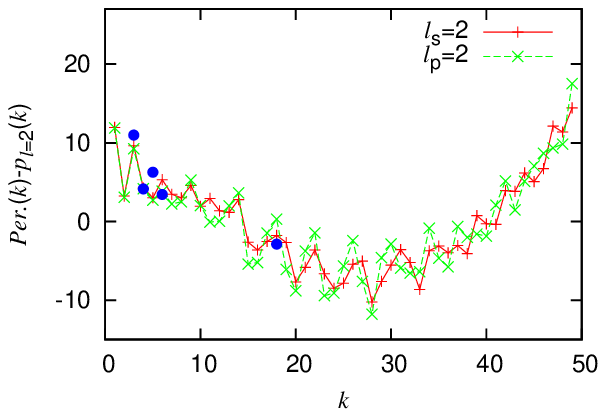}
\end{center}
\caption{Same as Fig. 4, but for $l$ = 2 modes.}
\end{figure}

In Fig. 3, we show the compositions profiles and corresponding Brunt-V\"ais\"al\"a frequency for two models of $M_{*}$ = 0.545\,$M_{\odot}$, $T_{\rm eff}$ = 30200\,K, and log($M_{\rm He}/M_{*}$) = -6.0. The red solid line is the model evolved by adopting the screened Coulomb potential, while the green dashed one is the model evolved by adopting the pure Coulomb potential. The differences locate at the interface area of C/O-He, the area of log($M_{\rm r}/M_{*}$) around -6.0. Correspondingly, there are some differences for the Brunt-V\"ais\"al\"a frequency at the area of log($M_{\rm r}/M_{*}$) around -6.0 in the up panel. Modes with high $k$ values, who are sensitive to this area, have a chance to distinguish the differences.

In Fig. 4 and Fig. 5, we show the fitting results for $l$ = 1 modes and $l$ = 2 modes based on the two models calculated in Fig. 3. A linear fit of the calculated periods in Table 4 were performed and subtracted from corresponding periods. Then, Fig. 4 and 5 can be drawn. For $l$ = 1 modes, the value of $k$ is 1, 2, 3, 4, 5, and 6. For $l$ = 2 modes, the value of $k$ is 3, 4, 5, 6, and 18. Except for the mode of $l$ = 2, $k$ = 18 ($f_{9}$), the other 10 observed modes are short period modes with low $k$ values. The calculated corresponding low $k$ modes are basically overlapping in Fig. 4 and 5. The low $k$ modes are not sensitive to the thin area of log($M_{\rm r}/M_{*}$) around -6.0. Actually, there are only small differences between the models with the screened potential (red, plus signs) and the pure potential (green, x's). However, mode $l$=2, $k$=18 is very sensitive to this layer and predicted periods are very different between the two models. This mode is much better fitted by the model with the screened potential.

In general, most of the observed modes for PG 0112+104 are low $k$ modes, which are not sensitive to the interface area of C/O-He. The differences are not obvious for the DBV star models taking element diffusion effect with pure Coulomb potential and screened Coulomb potential fitting PG 0112+104. If we observe many high $k$ modes, the differences will be obvious.

\section{Discussion and conclusions}

In this paper, we reported the results for the application of screened Coulomb potential in fitting DBV star PG 0112+104. PG 0112+104 was observed for 78.7 days from 2016 January 04 during Campaign 8 of the $Kepler$ 2 mission. Based on the observations, Hermes et al. (2017) made a detailed mode identification for PG 0112+104. We briefly review the photometric observations and mode identifications. There are 5 $l$ = 1 modes, 3 $l$ = 2 modes, and 3 $l$ = 1 or 2 modes identified, as shown in Table 2. Then, we followed the evolution of white dwarfs down their cooling tracks. Changes in their chemical composition profiles due to diffusion were modelled using two different descriptions of the Coulomb potential. Unlike the models of Hermes et al. (2017), the DBV star models in this paper have evolved core compositions (from $MS$ stage to the $WD$ cooling stage) and element diffusion profiles with pure and screened Coulomb potential. The eigen-modes are calculated by the pulsation code of Li (1992a,b) and then are used to fit the identified modes of PG 0112+104. Fitting PG 0112+104, we studied the differences for the DBV star models taking element diffusion effect with pure Coulomb potential and screened Coulomb potential.

For our fitting results, $T_{\rm eff}$ is 1.3\% smaller than results of Hermes et al. (2017) and $M_{\rm *}$ is 0.9\% larger than their results. However, the fitting error $\sigma_{RMS}$ is reduced by 27\%, as shown in Table 5. The asteroseismological studies are basically consistent with previous spectroscopic studies for PG 0112+104 (Dufour et al. 2010, Bergeron et al. 2011).

Fitting PG 0112+104, the differences are not obvious for the DBV star models taking element diffusion effect with pure Coulomb potential and screened Coulomb potential, as shown in Fig. 1 and 2. The differences are very obvious for the composition profiles of the C/O-He interface for the two scenarios. The high $k$ modes are very sensitive to the area of C/O-He interface. However, all the observed modes for PG 0112+104 except for the 503\,s mode are low $k$ modes. They are not sensitive to the area of C/O-He interface. Therefore, the overall improvement of the $\sigma_{RMS}$ is only 4\% for the screened Coulomb potential comparing with the pure Coulomb potential. However, the one high $k$ mode observed in PG 0112+104 is much better reproduced by the model calculated with the screened potential. If there were high quality photometric observations with many high $k$ modes for some DBV star, the differences would be detected.

\section{Acknowledgements}

The work is supported by the NSFC of China (Grant No. 11563001) and the Yunnan Applied Basic Research Project (2015FD044). Thank the Open Research Program of key Laboratory for the Structure and Evolution of Celestial Objects, Chinese Academy of Sciences (OP201502) and the Research Fund of Chuxiong Normal University (XJGG1501). We are very grateful to Q. S. Zhang, Y. Li, T. Wu, and X. H. Chen for their kind suggestions.

\label{lastpage}

\end{document}